\documentclass[%
 reprint,
 amsmath,amssymb,
 aps,
 pra,
floatfix
]{revtex4-2}

\usepackage[utf8]{inputenc}
\usepackage{graphicx}
\usepackage{dcolumn}
\usepackage{bm}
\usepackage{physics}
\usepackage{wrapfig}
\usepackage{xcolor}
\usepackage[normalem]{ulem}

\begin{document}

\title{Topological entanglement stabilization in superconducting quantum circuits}

\begin{abstract}
Topological properties of quantum systems are one of the most intriguing emerging phenomena in condensed matter physics. A crucial property of topological systems is the symmetry-protected robustness towards local noise. Experiments have demonstrated topological phases of matter in various quantum systems. However, using the robustness of such modes to stabilize quantum correlations is still a highly sought-after milestone. In this work, we put forward a concept of using topological modes to stabilize fully entangled quantum states, and we demonstrate the stability of the entanglement with respect to parameter fluctuations. Specifically, we see that entanglement remains stable against parameter fluctuations in the topologically non-trivial regime, while entanglement in the trivial regime is highly susceptible to local noise. We supplement our scheme with an experimentally realistic and detailed proposal based on coupled superconducting resonators and qubits. Our proposal sets a novel approach for generating long-lived quantum modes with robustness towards disorder in the circuit parameters via a bottom-up experimental approach relying on easy-to-engineer building blocks.
\end{abstract}

\author{Guliuxin Jin}
\email{g.jin@tudelft.nl}
\affiliation{Kavli Institute of Nanoscience, Delft University of Technology, 2628 CJ Delft, the Netherlands}%
\author{Eliska Greplova}%
\email{e.greplova@tudelft.nl}
\affiliation{Kavli Institute of Nanoscience, Delft University of Technology, 2628 CJ Delft, the Netherlands}%

\maketitle

\section{Introduction}

Topological quantum states have been an object of great interest to physicists over the last few decades due to their curious properties~\cite{wang2017topological, acin2018quantum, keimer2017physics}. Starting with the discovery of the quantized conductivity in the quantum Hall effect~\cite{thouless1982quantized}, the field has developed a series of exciting discoveries in modern topics like topological insulators~\cite{kane2005quantum, bernevig2006quantum}, Weyl semimetals~\cite{wan2011topological, soluyanov2015type}, and topological superconductors~\cite{qi2009time,fu2010odd,sasaki2011topological,qi2010chiral}. At the same time, topological phenomena offer promising applications. For example, non-interacting topological models, represented by the band theory, give rise to the field of spintronics~\cite{vzutic2004spintronics}, while certain interacting models are predicted to lead to ground-breaking new applications such as topological quantum computing~\cite{aasen2016milestones}. 

A great success of modern condensed matter theory is the discovery of that topological phases of matters are generally subject to topological invariants related to the global symmetries~\cite{bernevig2013topological}. These invariants were originally connected to the phases of quantum systems. Still, recent studies have shown that a number of topological phenomena initially  observed in non-interacting quantum systems are reproducible in purely classical systems~\cite{susstrunk2015observation, susstrunk2016classification, kane2014topological, paulose2015topological, chen2014nonlinear, chen2016topological,nash2015topological, serra2018observation, imhof2018topolectrical,kollar2019hyperbolic,kim2021quantum}. Recently the straightforward engineering principles discovered through the field of classical metamaterials have been adopted to  quantum systems, where such metamaterials were then reproduced~\cite{kollar2019hyperbolic,kim2021quantum}. In this work, we put forward a new concept of topology-stabilized quantum entanglement, inspired by classical topological metamaterials, and thus introduce a bottom-up method to engineer topological  metamaterials inherently manifesting quantum properties. Specifically, we propose a method to generate long-range entangled states of topological modes in a one-dimensional (1D) system and provide a detailed analysis of the noise robustness such modes possess.

This manuscript is organized as follows:
In Sec.~\ref{sec:sshmodel} we give a brief overview of the Su-Schrieffer-Heeger (SSH) model, a basic building bloc of our proposal. In Sec.~\ref{sec:entanglingthetopologicalmodes} we propose a sequence to entangle two spatially separated topological edge modes. In Sec.~\ref{sec:entanglementstabilityanalysis} we statistically analyze the stability of such topological entanglement in the presence of parameter fluctuations. In Sec.~\ref{sec:targetingthemaximumentanglement} we discuss practical methods to address topological entanglement in disordered systems. Finally, in Sec.~\ref{sec:conclusionandoutlook} we give a conclusion of our study and discuss the outlook.

\section{Su-Schrieffer-Heeger Model}
\label{sec:sshmodel}

\begin{figure}[t]
\centering
\includegraphics[width= 0.95\linewidth]{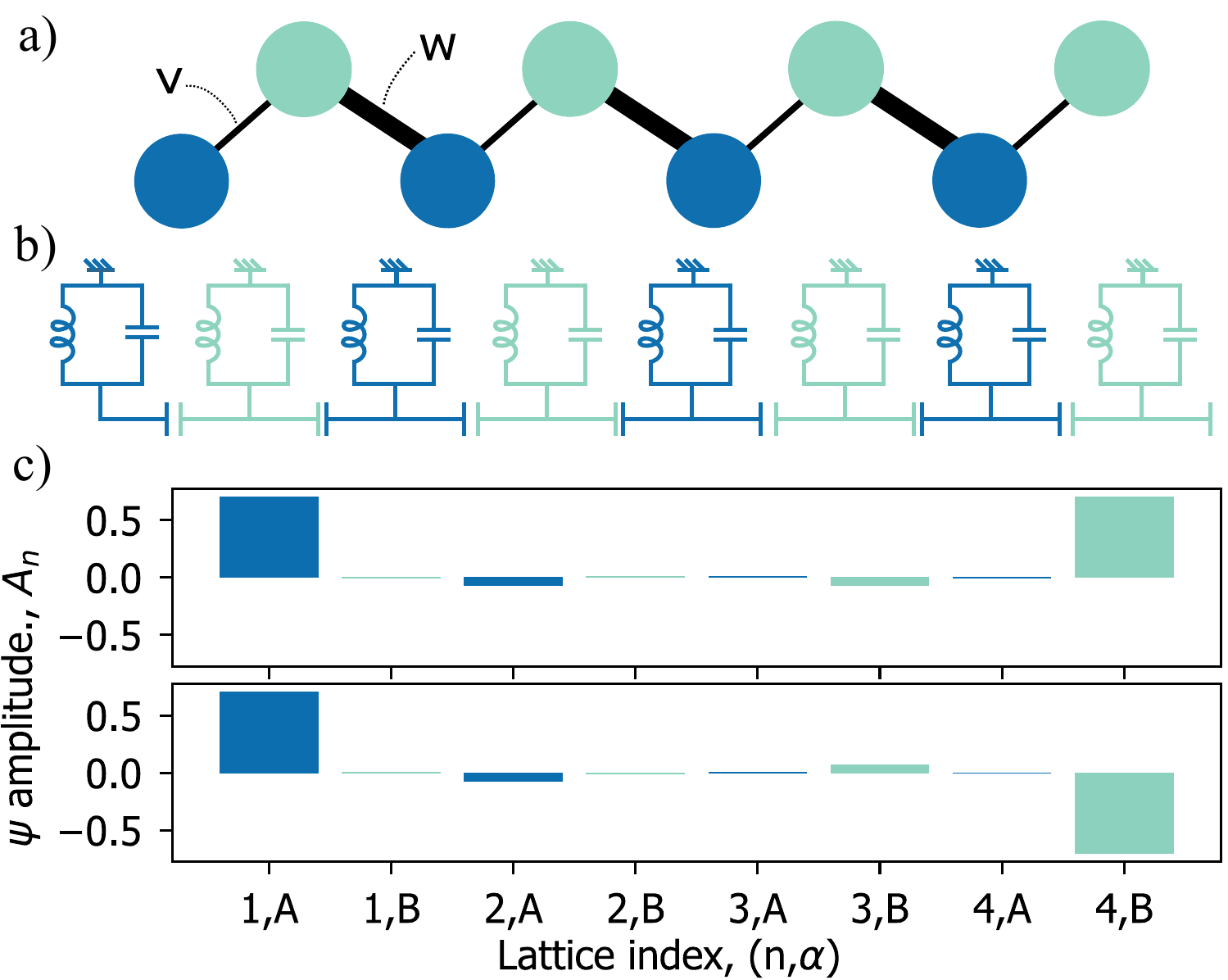}
\caption{a) Schematic plot of the SSH model in a finite system. The blue and green dots are type-$A$ and $B$ sublattice sites. The thin and thick lines represent the intra-cell coupling $v$ and inter-cell coupling $w$ respectively. b) Experimental realization with capacitively coupled superconducting resonators. c) Wave-function $\psi$ amplitude $A_n$ of the two mid gap zero-energy edge modes with $v=0.1, w = 1.0$. Upper panel: Plot of the symmetric edge mode; lower panel: Plot of the anti-symmetric edge mode.}
\label{fig:ssh_edgemode_combi}
\end{figure}

The Su-Schrieffer-Heeger (SSH) model~\cite{su1979solitons} is a prominent simple non-interacting model exhibiting topological properties. It was proposed to describe spinless fermions hopping on a one-dimensional lattice with staggered hopping amplitudes~\cite{su1979solitons}, see Fig.~\ref{fig:ssh_edgemode_combi}(a). This model leads to topological edge modes supported on the ends of a lattice in the topologically non-trivial phase, see Fig.~\ref{fig:ssh_edgemode_combi}(c). While originally proposed for fermions, the model can readily be implemented as an array of superconducting resonators~\cite{kim2021quantum}, see Fig.~\ref{fig:ssh_edgemode_combi}(b).

The SSH Hamiltonian~\cite{su1979solitons} is given by
\begin{equation}\label{eq:ssh_eq}
    H=\sum_{n=1}^{N}(v_n c^{\dagger}_{n,A}c_{n,B}+ w_{n}c^{\dagger}_{n,B}c_{n+1,A}+h.c.),
\end{equation}
where $c^{\dagger}_{n,\alpha}$ ($c_{n,\alpha}$) is the creation (annihilation) of a particle on lattice site ($n,\alpha$) with unit cell index $n \in [1,N]$ and sublattice index $\alpha \in {A,B}$. The intra-cell to inter-cell hopping ratio, $v/w$, controls the topological phase transition. The model is in the topologically non-trivial phase at $v/w < 1$ where two  mid-gap states exist. Such states are exponentially localized at the edge lattices, as shown in Fig.~\ref{fig:ssh_edgemode_combi}(c). The two edge modes are robust towards arbitrary disorder as long as the disorder respects the chiral symmetry of the Hamiltonian. The energy gap closes at $v/w = 1$, which indicates the topological phase transition. When $v/w > 1$, the model is in the trivial phase where two mid-gap edge states merge into the bulk band, and the edge localization is no longer valid. In the context of superconducting circuits, resonator transmission and reflection measurements can readily confirm the appearance of the mid-gap modes in the topological phase of the SSH model. We provide a detailed discussion of the SSH model and its spectrum in Appendix~\ref{app:ssh}. 


\section{Entangling the Topological Modes}
\label{sec:entanglingthetopologicalmodes}

\begin{figure}[b]
\centering
\includegraphics[width=1.0\linewidth]{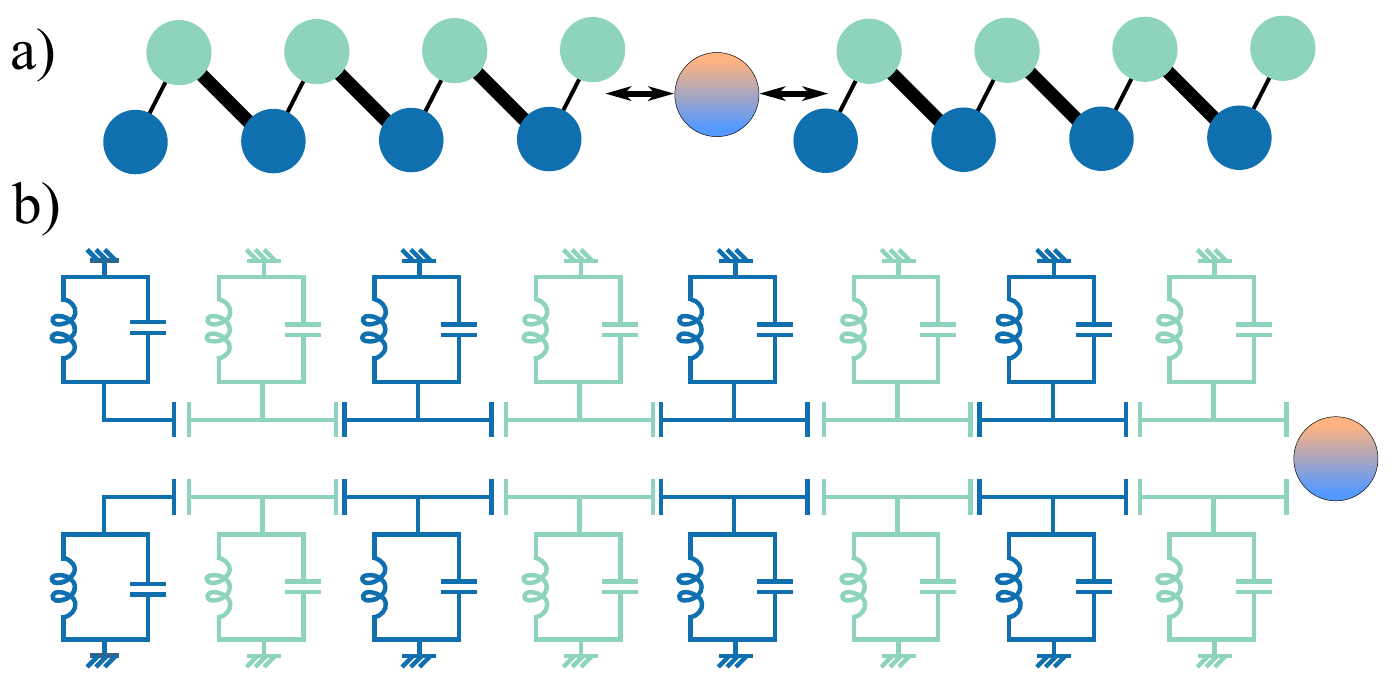}
\caption{a) Schematic plot of two SSH arrays coupled to one qubit - each chain consisting of four unit cells of two sublattice sites. Blue and green colors represent the $A$ and $B$ types of sublattice. Orange-blue gradient circle represents the qubit. Arrows represent dispersive coupling between the qubit and the edge lattice sites. b). Each sublattice site represents one superconducting LC resonator. The varying capacitive coupling between resonators is indicated with the capacitor distance.}
\label{fig:ssh_qubit}
\end{figure}

In the context of quantum entanglement, we propose a scheme to entangle two spatially separated topological SSH edge modes and investigate the robustness of such entanglement against disorders. This task can be accomplished by engineering a system consisting of a single qubit and two 1D SSH arrays, see Fig.~\ref{fig:ssh_qubit}(a). In Fig.~\ref{fig:ssh_qubit}(b) we show how this architecture can be engineered with capacitively coupled superconducting (SC) resonators arrays~\cite{blais2004cavity,koch2007charge,wendin2017quantum,frey2012dipole}. The specific realization of the qubit is flexible and platform-dependent. It can be achieved, for example, by a transmon within a purely superconducting setting~\cite{koch2007charge, houck2008controlling, scigliuzzo2021extensible, wendin2017quantum} or by a quantum dot in the case of hybrid devices \cite{scarlino2019all, frey2012dipole, de2010hybrid, delbecq2011coupling, hensgens2017quantum, nowack2007coherent, hendrickx2021four}.

When both SSH chains are in the topologically non-trivial phase, i.e., inter-cell coupling $w$ is larger than intra-cell coupling $v$, we can observe two zero-energy modes localize at the edges in each array. The system in Fig.~\ref{fig:ssh_qubit}(a) is described, in a frame rotating at the resonator and qubit frequencies, by the Hamiltonian
\begin{equation}\label{eq:H_full2}
\begin{split}
    H =  \sum_{<i,j>} J_{1,ij} a_{1,i}^{\dagger} a_{1,j} + 
    \sum_{<i,j>} J_{2,ij} a_{2,i}^{\dagger} a_{2,j}\\
    +    \xi_1 \sigma_z a_{1,N}^{\dagger} a_{1,N}+
    \xi_2 \sigma_z a_{2,1}^{\dagger} a_{2,1},
\end{split}
\end{equation}
where $a_{X,i}^{\dagger}$ ( $a_{X,i}$) are creation (annihilation) operators in the SSH chains. $X = 1,2 $ represents the two SSH chains and $i$ is the unit cell within a chain. In this setting, the modes excited by driving of the resonator array are bosons. The qubit is described by the operator $\sigma_z = \ket{e}\bra{e} - \ket{g}\bra{g}$. We assume that the qubit is coupled to the resonators in the dispersive regime with $\xi$ the qubit state-dependent dispersive frequency-shift of the resonators, given by $\xi_X = g^2 / \delta_X$, where $g$ is the dipole-coupling between the resonator and the qubit. This assumption is valid when a large qubit-resonator detuning $\delta_X$ with respect to the coupling strength $g$ is present~\cite{koch2007charge,blais2004cavity}. 

We start investigating the entanglement in the resonator-qubit system based on Hamiltonian Eq.~\eqref{eq:H_full2}. A tripartite quantum state is created by exciting this system to a state that couples the qubit to the topological edge modes in both resonator arrays. By three parties, we here refer to topological edge modes in first resonator array, the qubit, and the topological edge modes in the second resonator array. Below we show that the qubit projective measurement results in a maximally entangled state among the topological modes on SSH arrays, whose robustness we wish to analyze.

Numerical analysis of systems shown in Fig.~\ref{fig:ssh_qubit} is done by the exact diagonalization, where the single particle basis of each SSH chain is deployed to numerically diagonalize the Hamiltonian Eq.~\eqref{eq:H_full2}. Note that the non-interacting nature of the bosonic excitations does not restrict each SSH chain to only hosting a single excitation, it merely means we are expressing our wave-function in the basis of single-particle excitations. The exact diagonalization gives 128 eigenstates of the Hamiltonian Eq.~\eqref{eq:H_full2}.

We proceed by performing projective measurement in the $\sigma_z$ basis of the qubit, thereby projecting the qubit state either to the ground ($\ket{g}$) or excited ($\ket{e}$) state. The probability to measure the $\ket{e}$ and $\ket{g}$ states are equal. The details of the calculation are in App.~\ref{app:ednegativity}. The qubit projection yields number of possible entangled states between the remaining two SSH arrays. When the desired eigenstate of the system is excited followed by a projective measurement yielding the state $\ket{e}$, the density matrix of the two SSH arrays shows the appearance of maximal entanglement, see Fig~\ref{fig:density_plot}

\section{Entanglement stability analysis}
\label{sec:entanglementstabilityanalysis}

Negativity is a measure of bipartite quantum entanglement. It is derived from the positive partial transpose (PPT) criterion for separability of quantum states~\cite{zyczkowski1998volume,peres1996separability,horodecki2001separability} and is an entanglement monotone. Therefore it reaches its maximum on the maximally entangled states. Zero negativity value indicates no distillable entanglement, while non-zero value indicates the existence of distillable entanglement. We use negativity to characterize the entanglement generated in the two SSH system above.

Consider a general system with density matrix $\rho$ composed of two subsystems $A$ and $B$. The partial transposition of $\rho$ with respect to subsystem $B$ is given by~\cite{zyczkowski1998volume,peres1996separability,horodecki2001separability}
\begin{equation}
\label{eq:ptrho}
\begin{split}
    \rho ^{T_{B}}:= &(I\otimes T)(\rho ),
\end{split}
\end{equation}
which is the identity map applied to $A$  and the transposition map applied to $B$. Negativity can be calculated through the absolute sum of the negative eigenvalues of $\rho^{T_B}$~\cite{zyczkowski1998volume,peres1996separability,horodecki2001separability}, given by
\begin{equation}
\begin{split}
 {\mathcal {N}}(\rho )=\left|\sum _{\lambda _{i}<0}\lambda _{i}\right|=\sum _{i}{\frac {|\lambda _{i}|-\lambda _{i}}{2}},
\end{split}
\end{equation}
where $\lambda_i$ are all the eigenvalues of $\rho^{T_B}$. Note that $\mathcal{N} \in [0, 0.5]$ for bipartite entanglement. Maximal entanglement states, such as Bell states, reach the value $\mathcal{N} = 0.5$. The reconstruction of the density matrix $\rho$ from experimental data is dependent on the specific implementation. The quantum state tomography in the context of superconducting qubits was shown in Refs.~\cite{besse2020realizing, eichler2012characterizing}.

\begin{figure}
\centering
\includegraphics[width= 1.0\linewidth]{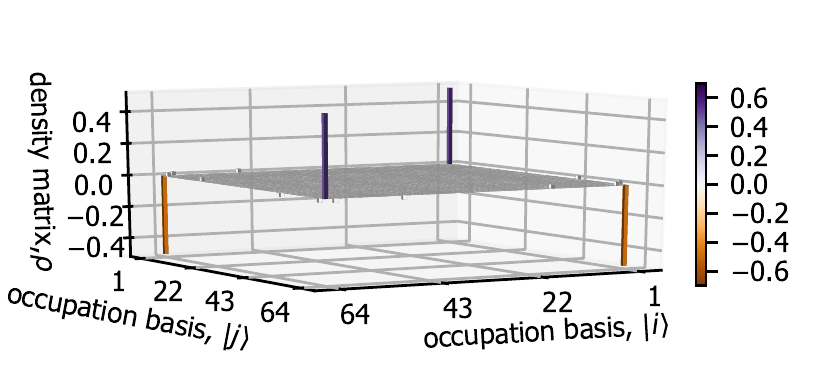}
\caption{Density matrix $\rho$ of a chosen entangled states after the projective qubit measurement. The chosen state is a eigenstate of Eq.~\eqref{eq:H_full2} projected to qubit $\ket{e}$. The parameters are chosen to be: inter-cell coupling $w = 10$, intra-cell $v = 0.5$, dispersive coupling rate $\xi_1 = \xi_2 = 0.001$. The occupation basis $\ket{1}$ and $\ket{64}$ correspond to single particle excitation at the edges in both SSH1 and SSH2. }
\label{fig:density_plot}
\end{figure}

By inspecting all the eigenstates of Hamiltonian Eq.~\eqref{eq:H_full2}, we find that the proposed system indeed hosts states that lead to entanglement between two SSH chains. For example, in Fig.~\ref{fig:density_plot} we show the density matrix $\rho$ of a $\mathcal{N} = 0.5$ state hosted by the two SSH chains after the qubit has been measured and projected onto the state $\ket{e}$ (here both SSH chains are in the topological regime). From the structure of the density matrix it is visible that the amplitudes are located at the edges of the two arrays according to the definition of the occupational basis. We emphasize that in the two-SSH system, each chain acts as a subsystem defined in Eq.~\eqref{eq:ptrho}. Therefore, the entanglement measured by the negativity is indeed between two subsystems, i.e., two SSH chains. 

\begin{figure}[t]
\centering
\includegraphics[width= 1.0\linewidth]{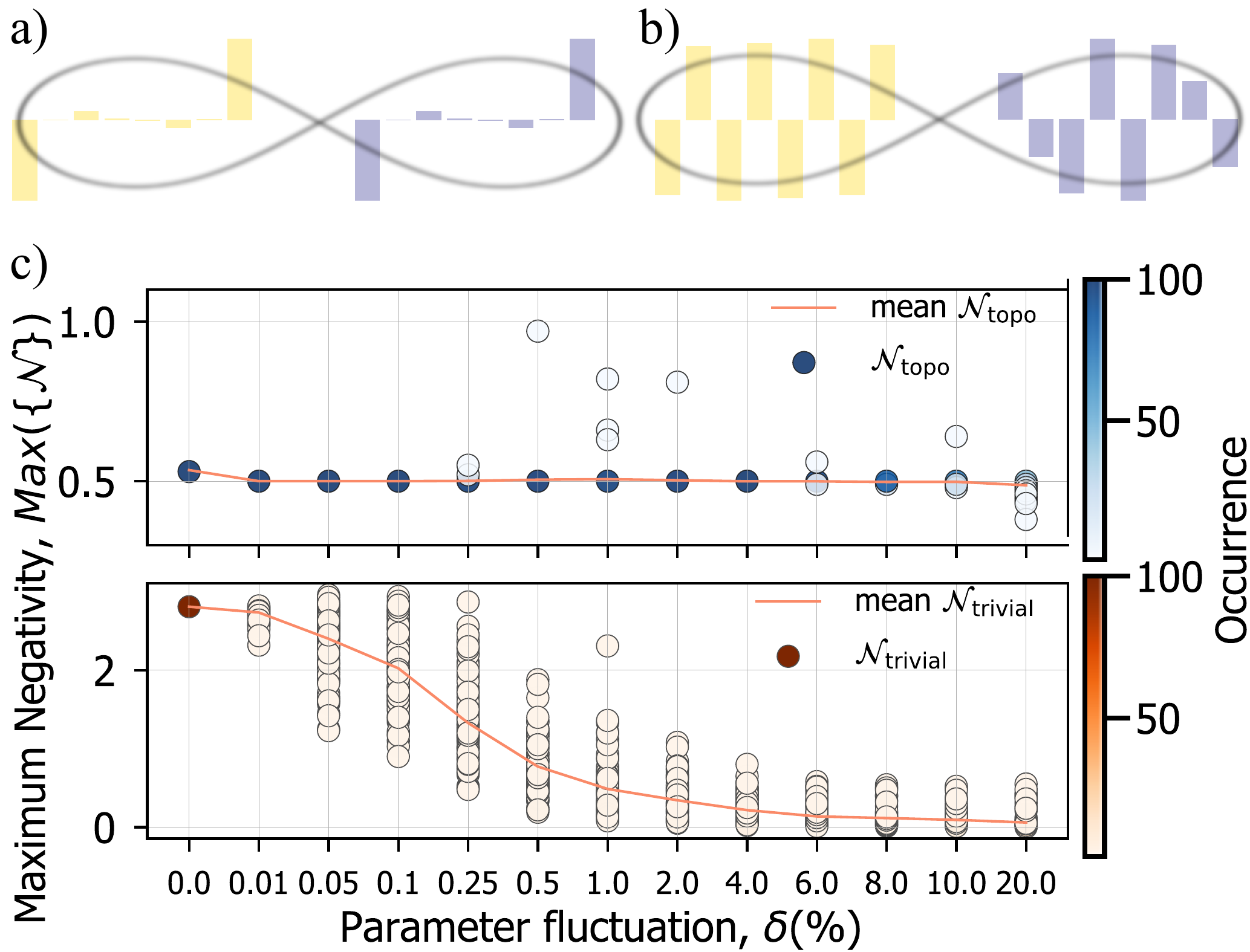}
\caption{
a) Schematic visualization of entanglement between two SSH chains in the topologically non-trivial regime. The state is chosen from a random Hamiltonian of parameter fluctuation $\delta = 0.1\%$ with $v=0.1, w=1$. 
b) Schematic visualization of entanglement in the trivial regime. The state is chosen from a random Hamiltonian of parameter fluctuation $\delta = 0.1\%$ with $v=1, w=0.1$.
c) Maximal negativity $\textit{Max}(\{\mathcal{N}\})$ as a function of parameter fluctuation strength $\delta$ in both topological and trivial regimes. At each parameter fluctuation strength $\delta$, the maximal negativities for each random Hamiltonian are shown. 100 Hamiltonians are sampled at each $\delta$. The color gradient indicates the number of occurrences per specific value of negativity. The upper panel shows the SSH chains in topological regime with $v=0.1, w=1$, while the lower panel is in trivial regime with $v=1, w=0.1$. The mean negativity over all random Hamiltonians is indicated with the orange line.}
\label{fig:neg_disorder_sample_combi}
\end{figure}

However, selecting the exact eigenstate that leads to maximal entanglement can be experimentally challenging because of the complicated spectrum under disorders. Let alone that noise can lead to decoherence of the entanglement itself.  We will now analyze the selection feasibility when many nearly degenerate eigenstates are present, particularly the existence of entanglement states when disorder is presented.

In our entanglement analysis, the first step is to investigate the robustness of the entanglement against parameter fluctuation disorders $\delta$. Now consider the case where the hopping parameters $v, w$ have certain degrees of uncertainty which we numerically simulate by the random distributions. The Hamiltonian Eq.~\eqref{eq:H_full2} is prepared such that both SSH arrays are in the topological phase with fluctuating $v, w$. Specifically, the Hamiltonian hopping amplitudes, given by value $v (w)$, can fluctuate uniformly within the interval $[v-\delta/2,v+\delta/2 ] $ $([w-\delta/2,w+\delta/2 ])$. The exact diagonalization of a randomly sampled Hamiltonian yields its 128 eigenstates. For each eigenstate, we project out the qubit and evaluate the negativity $\mathcal{N}$ on the remaining two-SSH system.

Next, we identify the state with the highest negativity among all the eigenstates. In the topological regime, we find out that such highly entangled states occupy the edges of each chain, the wave function of which is schematically shown in Fig.~\ref{fig:neg_disorder_sample_combi}(a). In contrast, when a Hamiltonian is in the trivial regime, the maximally entangled state is distributed across the whole array, as shown in Fig.~\ref{fig:neg_disorder_sample_combi}(b). Thus, we anticipate that the entanglement in the topological regime will remain stable against local parameter fluctuations since it shares the profile feature similar to the edge modes in the SSH model. On the other hand, the maximally entangled modes in the trivial regime is expected to be strongly susceptible to fluctuations. We quantitatively assess the anticipated robustness towards parameter fluctuations by calculating maximal negativity generated from $100$ random Hamiltonians for each value of parameter fluctuation, $\delta$. In each Hamiltonian, all 128 eigenstates are projected to $\ket{e}$ qubit state without loss of generality, and the corresponding 128 negativities, $\{ \mathcal{N} \}$, are calculated. We then evaluate the maximum negativity value. While we are displaying the results of the $\ket{e}$ state projection, it is worth noting that projecting onto the $\ket{g}$ state yields similar results.

In the topological regime (upper panel of Fig.~\ref{fig:neg_disorder_sample_combi}(c)) the maximal negativity $\textit{Max}(\{\mathcal{N}\})$ distribution is centered around $\mathcal{N} = 0.5$. Additionally, this value remains robust with respect to increased parameter fluctuation $\delta$. In the trivial regime (lower panel of Fig.~\ref{fig:neg_disorder_sample_combi}(c)) we observe that the maximal negativity $\textit{Max}(\{\mathcal{N}\})$ distribution changes rapidly with the parameter fluctuation and eventually decreases to zero. More importantly, the maximal negativity $\textit{Max}(\{\mathcal{N}\})$ shows a large spread over the sampled Hamiltonians, for example, the $\textit{Max}(\{\mathcal{N}\})$ distribution at $\delta = 0.25\%$ in the lower panel of Fig.~\ref{fig:neg_disorder_sample_combi}(c). 
In the trivial regime at the fixed parameter fluctuation, the fact that the maximal negativity varies for different samples indicates that after performing the measurement, the states collapse to random states in the Hilbert space, and the structure of their correlations does not remain fixed. The resulting entanglement structure is discussed in App.~\ref{app:othertypeentanglements}. We have also observed that the maximal negativities $\textit{Max}(\{\mathcal{N}\})$ in the trivial regime exceed the $\mathcal{N} = 0.5 $ limit. This behavior indicates that such entanglement likely goes beyond the bipartite entanglement. We discuss this phenomenon in detail in App.~\ref{app:othertypeentanglements}.

In summary, we observe that the existence of a maximally entangled state is highly robust when the two SSH chains are in the topological regime, while the entangled states has random negativity value and are prone to disorder in the trivial regime. From now on, we consider the $\mathcal{N} = 0.5$ states in the topological regime as the target state.


\section{Targeting the maximum entanglement}
\label{sec:targetingthemaximumentanglement}
\begin{figure}[t]  
\centering
\includegraphics[width= 1.0\linewidth]{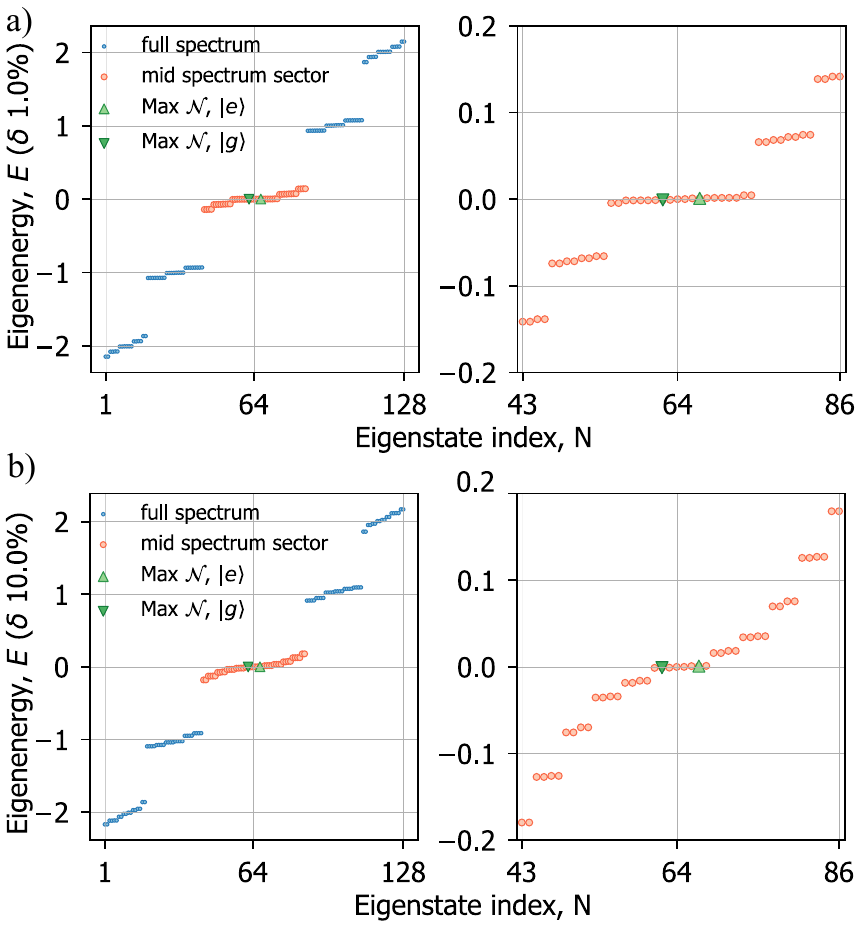}
\caption{a) Eigenenergy spectrum with $\delta = 1\%$. b) Eigenenergy spectrum with $\delta =  10\%$. Both  are in the topological regime ($ v = 0.1, w = 1.0$). The two green upward and downward triangles represent the two maximally entangled states when the qubit is projected to $\ket{e}$ or $\ket{g}$, respectively. The middle near-zero-energy section in the spectrum (highlighted in orange) is enlarged in the right panel.}
\label{fig:spectrum_combi}
\end{figure}

In order to properly address the target state, such as in Fig.~\ref{fig:neg_disorder_sample_combi}(a), we first study the energy spectrum of the proposed system and the spectrum change as a function of hopping parameter fluctuations. We show two eigenenergy spectra of Eq.~\eqref{eq:H_full2} with parameter fluctuation $\delta = 1.0\% $ and $\delta = 10.0\% $ in the topological regime ($v = 0.1, w=1.0$) in Fig.~\ref{fig:spectrum_combi}. The eigenstates are sorted by their energies and we refer to the sorted eigenstates by their eigenstate index (x-axes in Fig.~\ref{fig:spectrum_combi}). From our previous study where we calculate the negativity $\mathcal{N}$ of each eigenstate after qubit projection, we find out that the target states with maximal entanglement $\mathcal{N} = 0.5$ exist in the near zero energy section (see triangles in Fig.~\ref{fig:spectrum_combi}). Our study also finds that the increased parameter fluctuation $\delta$ yields a broader splitting between each eigenstate in energy spectrum, see the right panels in Fig.~\ref{fig:spectrum_combi}(a) and (b). This means that a certain amount of parameter fluctuation lifts the degeneracy and makes it easier to address certain states in the frequency space. This fact indicates the experimental applicability of our scheme.

\begin{figure}[t]  
\centering
\includegraphics[width= 1.0\linewidth]{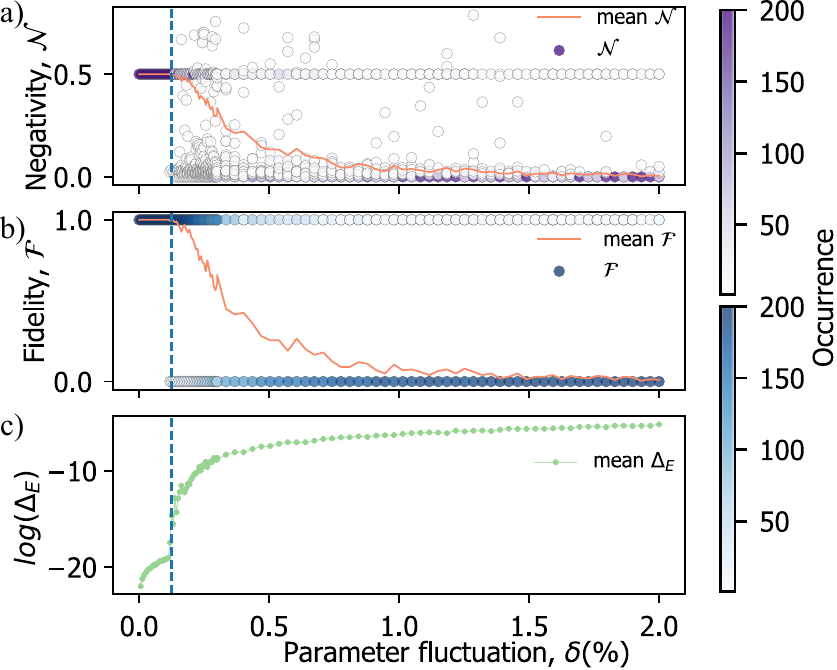}
\caption{Tracking the state $\ket{73}$ across the parameter fluctuation $\delta$.
a) Negativity of state $\ket{73}$ projected to $\ket{e}$.
b) Fidelity between the disordered $\ket{73}$ with respect to state $\ket{73}_{\delta = 0}$.
c) Logarithm energy difference between disordered $\ket{73}$ and state $\ket{73}_{\delta = 0}$.
Each dot represents a random Hamiltonian.
At each parameter fluctuation $\delta$, 200 random Hamiltonians are generated. The color gradient shows the occurrence of negativity/fidelity at each value, i.e., the overlapping Hamiltonians. The Hamiltonians are in the topological regime with $v =0.1, w=1$. The vertical dashed lines indicate that the system undergoes a cross-over at a parameter fluctuation around $\delta_t = 0.125\%$.}
\label{fig:eigen_index_three_panels}
\end{figure}

Our simulations show that the two states yielding maximum bipartite entanglement (green triangles) appear at the fixed eigenstate index of the non-disordered Hamiltonian. Thus we now investigate the eigenstate index stability to address the target maximum entanglement state. For instance, the non-disordered topological Hamiltonian of $\delta = 0.0\%$ shows maximal entanglement at eigenstate index $73$ in the spectrum.

To keep the notation concise, we now refer to the eigenstate with index 73 as $\ket{73}$. Without loss of generality, we choose $\ket{73}$ to represent the $\mathcal{N} = 1/2$ topological entanglement state and analyze the stability of a specific eigenstate against parameter fluctuation. We trace this state starting from zero $\delta$ ($\delta = 0.0\%$) towards large $\delta$ and calculate three quantities: the negativity, the fidelity between  $\ket{73}$ at disorder ($\delta >0$) and $\ket{73}_{\delta=0}$, and the energy difference between $\ket{73}$ at disorder ($\delta >0$) and $\ket{73}_{\delta=0}$, see Fig.~\ref{fig:eigen_index_three_panels}. Generally, the fidelity between two quantum states, $\ket{\psi_a}$ and $\ket{\psi_b}$, is evaluated as
\begin{equation}\label{eq:fidelity}
    \mathcal{F}(a ,b )=|\langle \psi _{a }|\psi _{b }\rangle |^{2}.
\end{equation}

Since a eigenstate can be referred and therefore addressed by its eigenstate index, we statistically analyze the eigenstate index stability, we choose a parameter fluctuation $\delta$ grid evenly from $[0.0\% , 2.0\%]$. For each $\delta$, we initialize 100 random Hamiltonians in the topological regime: $v = 0.1, w = 1.0$. As shown in Fig.~\ref{fig:eigen_index_three_panels}, when parameter fluctuation $\delta$ grows beyond a certain threshold value (dashed vertical lines), the negativity distribution of $\ket{73}$ is no longer consistently at $0.5$ and the mean value $\overline{\mathcal{N}}$ decreases. Similarly the fidelity $\mathcal{F}$ between $\ket{73}$ and $\ket{73}_{\delta=0}$ deviates from $\mathcal{F} = 1$ when passing the threshold $\delta$. Moreover we also see a drastic change in the $\log(\Delta E)$ at the same threshold $\delta$. Note that the exact value of dashed vertical lines ($ \approx 0.125\%$ ) is only to show the instability of the target state eigenindex. We can conclude from these studies that although the existence of a state that leads to a negativity $\mathcal{N} = 0.5$ is validated by Fig.~\ref{fig:neg_disorder_sample_combi}, such states cannot be excited by simply addressing the specific eigenstate index from the eigenspectra. 

Despite the fact that the target topological entanglement state cannot be directly identified by the eigenindex, we observe that the state is generally located within a certain energy window.
Moreover, when a random disordered topological Hamiltonian has an eigenstate that leads to $\mathcal{N} = 0.5$ states, there will only be one such state. When parameter fluctuation $\delta$ changes, the energy of this target state is stable in comparison to the scale of the energy window width. Thus it is possible to locate the target state within such energy windows. The details of energy window preparation can be found in App.~\ref{app:energywindow}. In the case of $v = 0.1, w = 1.0$, the energy window has size $\Delta_E = 1.08\times 10^{-6}$ and centers around the target energy 
$E_{0,\ket{e}} = 9.9 \times 10^{-4}$ or symmetrically $E_{0,\ket{g}} = -9.9 \times 10^{-4}$. The energy window parameters will change for different qubit projections as well as for different $v/w$ ratios. The complete parameters are given in Tab.~\ref{tab:table1}.

Let us illustrate what these results would correspond to in the physical units in a typical superconducting experiment. Setting $w = 2\pi/50$MHz will result in the difference between $E_{0,\ket{e}}$ and $E_{0,\ket{g}}$ of $100$kHz and $\Delta$ of ca $50$Hz. Assuming typical resonator frequency of $5$GHz and Q-factor of $250 000$ we obtain a line-width of $20$kHz. The mid-gap states are thus individually addressable and the fluctuation of the energy window also does not pose an issue for this addressability. Additionally, the line-width could be much smaller by increasing the Q-factor which is routinely brought up to several million in the current experiments.

\begin{figure}[t]  
\includegraphics[width= 1.0\linewidth]{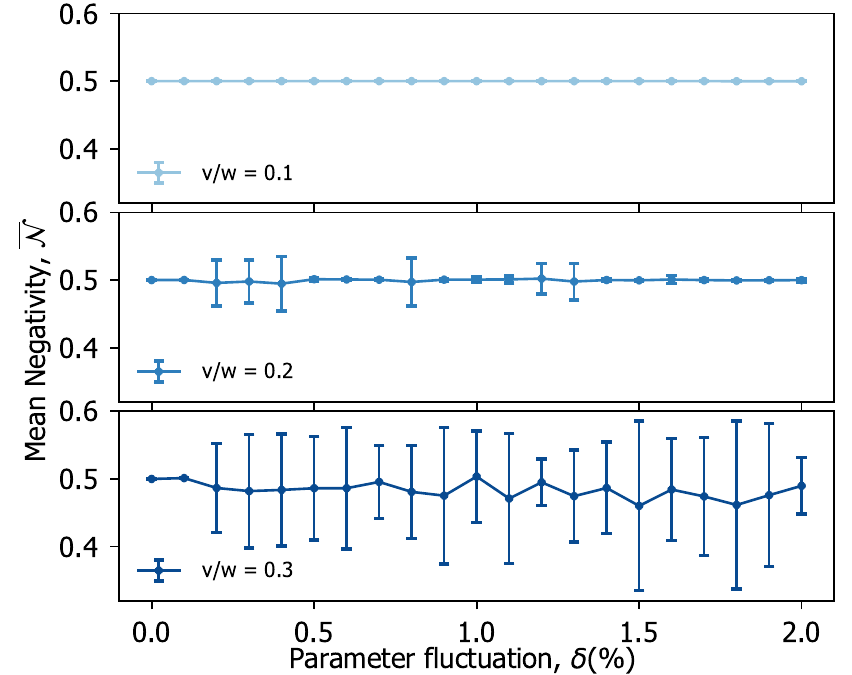}
\caption{Mean negativity $\overline{\mathcal{N}}$ as a function of parameter fluctuation strength $\delta$ under different $v/w$ ratios. The mean negativity is calculated from all states with eigenenergy in the energy window. Each point represents the mean negativity value with the standard deviation indicated with the error bar. Each data point is calculated from 100 randomly sampled Hamiltonian under the specific $\delta$.}
\label{fig:energy_window_mean_negativity}
\end{figure}

When we address all the eigenstates within the prescribed energy window, Figure~\ref{fig:energy_window_mean_negativity} shows the mean negativity $\overline{\mathcal{N}}$ and the corresponding standard deviation. If the energy window is empty, we will not enter the calculation of mean negatiivty. Each data point is calculated from 100 random Hamiltonians and averaged from both qubit projections. We find that the mean negativity $\overline{\mathcal{N}}$ is stable at small $v/w$ ratios as in the upper panel of Fig.~\ref{fig:energy_window_mean_negativity}. The negativity instability rises as the $v/w$ increases. The increased $\delta$ does not increase the instability. Thus, the non-trivial topology of the SSH model is critical in stabilizing the entanglement in the presence of disorders.

\begin{figure}[t]  
\centering
\includegraphics[width= 1.0\linewidth]{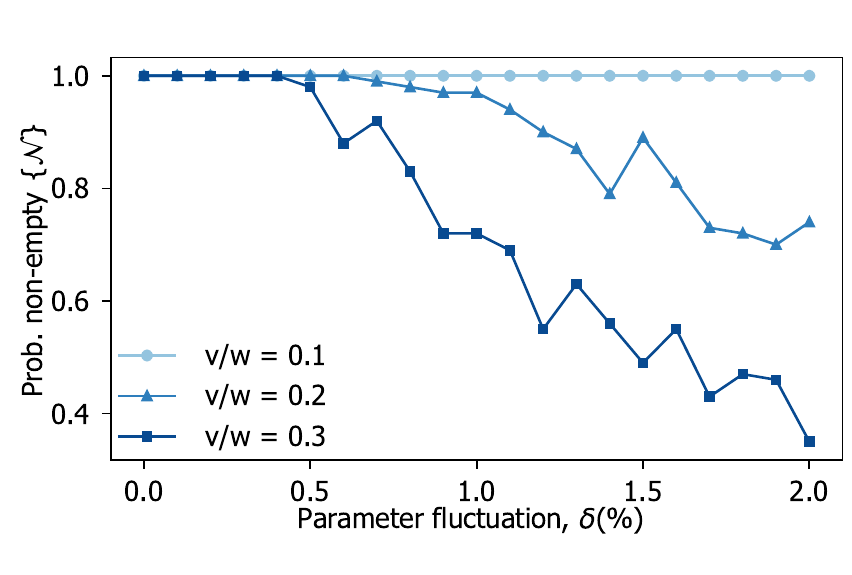}
\caption{The probability of the $v/w$ dependent energy window being non-empty in a Hamiltonian, at different parameter fluctuation $\delta$. The $v/w$ are chosen in consistence with Fig.~\ref{fig:energy_window_mean_negativity}. Each data point is calculated from 100 randomly sampled Hamiltonian under the specific $\delta$.}
\label{fig:energy_window_prob_non_zero}
\end{figure}

Besides the mean negativity $\overline{\mathcal{N}}$, we are also interested in the number of states within the energy window since the negativity is ill-defined in Fig.~\ref{fig:energy_window_mean_negativity} if the energy window contains zero state. Specifically, we study the probability of having an energy window being non-empty, i.e., the energy window addresses at least one state of arbitrary $\mathcal{N}$, at various $\delta $ and $v/w$. In Fig.~\ref{fig:energy_window_prob_non_zero}, each data point is averaged over 100 random Hamiltonians. We find that when deep in the topological regime ($v/w = 0.1$), the prescribed energy window will always hit a specific state, the negativity of which is shown in Fig.~\ref{fig:energy_window_mean_negativity}. In contrast, we find that the number of states within the target window decreases when the system moves away from deep topological regime, as shown in the case of $v/w = 0.2$ and $0.3$. The probability of having a non-empty window also decreases with $\delta $ in the latter two cases.  
Overall, the topological entanglement can be systematically achieved by addressing the specific energy window for a random Hamiltonian deep in the topological regime ($v/w = 0.1$). In this setup, the parameter fluctuation will not hinder the desired topological entanglement.


\section{Conclusion and Outlook}
\label{sec:conclusionandoutlook}
  
In this work, we proposed a protocol for realizing robust topological entanglement in superconducting circuits. Specifically, we theoretically construct an architecture containing SSH arrays and a single qubit. We showed that the Bell-like entanglement between topological edge modes could be achieved by projecting out the qubit. Such entanglement is robust against parameter fluctuations in the SSH resonator arrays. Additionally, we provided a detailed analysis of addressing the entanglement modes in the frequency space. We formulated a prescription on the rendered topologically stabilized entanglement as a function of parameter fluctuation. We conclude that there is always a unique way to prepare a maximally bipartite entanglement of topologically protected edge modes by targeting a specific frequency window.

By proposing an experimentally accessible scheme for the proof of the principle of topology stabilized entanglement, we put forward the concept of classical metamaterial-inspired  topological quantum devices engineering on firm quantum setups. This work can be used as a stepping stone for further topology stabilized quantum information processing and communication. The potential applications include more robust quantum communications links or robust on-chip entanglement.

The code needed to reproduce the results presented in this manuscript can be found in \cite{TopoEntanglementStabilization}.

\begin{acknowledgments}
We are thankful for enlightening discussions with Vincent Jouanny, Vera Weibel, Pasquale Scarlino. This publication is part of the project Engineered Topological Quantum Networks (with project number VI.Veni.212.278) of the research programme NWO Talent Programme Veni Science domain 2021 which is financed by the Dutch Research Council (NWO).
\end{acknowledgments}


\appendix
\section{SSH model spectrum}
\label{app:ssh}
The SSH model describes the non-interacting particle hopping on a one dimensional lattice composed of $N$ unit cells of sublattice $A$ and $B$. The lattice has staggered hopping amplitudes $v$ and $w$. Depending on the ratio of the two hoppings $v/w$, the SSH model has a topological phase transition. 

\begin{figure}[h]
\centering
\includegraphics[width= \linewidth]{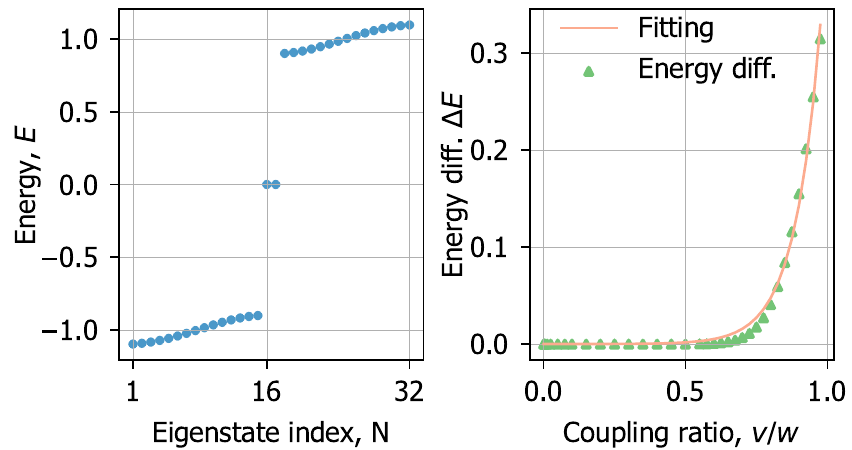}
\caption{Left: SSH energy spectrum  with $v/w = 0.5$ in the lattice of  16 lattice sites. Right: Energy difference between two mid gap edge modes, fitted with an exponential function.}
\label{fig:ssh_spectrum}
\end{figure}

Given the SSH Hamiltonian Eq.~\eqref{eq:ssh_eq}, we analyze the spectrum of this model in topologically non-trivial case with hoppings $v = 0.5, \ w = 1.0$. In a lattice of 16 lattice sites with open boundary condition, the energy spectrum is given in left panel of Fig.~\ref{fig:ssh_spectrum}. 
The two edge-states are shown in the middle of the gap. 
The two topological edge modes are degenerate in the periodic boundary condition, while the degeneracy is lifted in the finite systems with open boundary condition.
We find that the energy difference between two mid-gap edge states increases when we move away from the deep topological non-trivial regime, i.e., small $v/w$ values. Thus the degeneracy between two edge states is lifted when increasing $v/w$ in finite systems. The energy difference grows exponentially with respect to the $v/w$ ratio, as shown in the right panel of Fig.~\ref{fig:ssh_spectrum}. This also indicates that the gap begins to close exponentially.

Although originally proposed to describe fermions on a one dimensional lattice with staggered hopping, the experimental implementations of the SSH model can go beyond the fermionic systems. The photonic analog is realized in systems where superconducting qubits are coupled to a metamaterial waveguide \cite{kim2021quantum}. The phononic analog is demonstrated in mechanical metamaterials\cite{chen2014nonlinear}. More complicated 2D models with topological phases are also implemented using metamaterials \cite{owens2018quarter,susstrunk2015observation}. 

\section{Calculations on exact diagonalization and entanglement negativity}
\label{app:ednegativity}

In order to numerically analyze the Hamiltonian Eq.~\eqref{eq:H_full2} of the two SSH chains with a qubit, as shown in Fig.~\ref{fig:ssh_qubit}, we exact diagonalize the Hamiltonian in the single particle basis of each SSH chain  with the coupled two-level qubit system. The full multi-particle Hilbert space is given by the tensor product space of two groups of SSH single particle basis and the two-level qubit system:
\begin{equation}\label{eq:basis}
\begin{split}
    \{ \ket{b_{full}} \} = \{ \ket{b_{ssh1}} \otimes \ket{b_{qubit}} \otimes \ket{b_{ssh2}} \}.
\end{split}
\end{equation}
The single-particle basis of each SSH chain (4 unit cells with 2 sublattice sites on each) is given by
\begin{equation}
\label{eq:sshbasis}
\begin{split}
    \{&\ket{b_{sshX}}\} = \\
    &\{ \ket{1,0,0,0,0,0,0,0}, \ket{0,1,0,0,0,0,0,0},\\
    & \ket{0,0,1,0,0,0,0,0}, \ket{0,0,0,1,0,0,0,0},\\
    & \ket{0,0,0,0,1,0,0,0}, \ket{0,0,0,0,0,1,0,0},\\
    & \ket{0,0,0,0,0,0,1,0}, \ket{0,0,0,0,0,0,0,1}\},
\end{split}
\end{equation}
where each digit represents a lattice site among the 8 sublattice sites.  The qubit basis is given by
\begin{equation}
\begin{split}
    \{\ket{b_{qubit}}\} = 
    \{ \ket{e}, \ket{g}\}.
\end{split}
\end{equation}
Therefore the Hilbert space size considered is $8\times 8\times 2 = 128$.  Note that the single particle picture is deployed to describe the non-interacting particles in each SSH chain separately. In experimental implementation, we do not intend to restrict the entire circuit to have only single excitation.

The bipartite entanglement  between two SSH chains is achieved via the qubit projection, after preparing the entire coupled system in an eigenstate. The qubit projection operators are given by
\begin{equation}
\begin{split}
    \hat{P}_{\ket{e}} = \sum_{\ket{b_{qubit}} = \ket{e}} \ket{b_{full}}\bra{b_{full}},\\
    \hat{P}_{\ket{g}} = \sum_{\ket{b_{qubit}} = \ket{g}} \ket{b_{full}}\bra{b_{full}}.
\end{split}
\end{equation}
Each eigenstate can be projected to one of the two qubit states. The projection of 128 eigenstates onto both qubit states respectively yields 256 two-SSH states { with some states being identical to the others. The resulting Hilbert space of the two-SSH system has size 64. Here, the probability to measure the qubit state $\ket{e}$ is $50\%$, similarly for $\ket{g}$ the probability is also $50\%$.

After the qubit projection, the entanglement need to be measured by negativity. Here we elaborate the detail in the negativity calculation.
Consider a general system composed of two subsystems $A$ and $B$ with a total density matrix $\rho$. The total Hilbert space is ${\mathcal{H}}_{A}\otimes {\mathcal{H}}_{B}$. The density matrix $\rho$ is given by
\begin{equation}
    \rho =\sum _{{ijkl}}p_{{kl}}^{{ij}}|i\rangle \langle j|\otimes |k\rangle \langle l|.
\end{equation}
$\rho^{T_B}$ is the partial transpose of density matrix $\rho$ with respect to subsystem $B$, given by~\cite{zyczkowski1998volume,peres1996separability,horodecki2001separability}
\begin{equation}
\label{eq:pt}
\begin{split}
    \rho ^{T_{B}}:= &(I\otimes T)(\rho )=\sum _{ijkl}p_{kl}^{ij}|i\rangle \langle j|\otimes (|k\rangle \langle l|)^{T}\\
    =&\sum _{ijkl}p_{kl}^{ij}|i\rangle \langle j|\otimes |l\rangle \langle k|=\sum _{ijkl}p_{lk}^{ij}|i\rangle \langle j|\otimes |k\rangle \langle l|,
\end{split}
\end{equation}
where $(I\otimes T)(\rho )$ is the identity map applied to the A party and the transposition map applied to the B party. 

Negativity can be computed through the absolute sum of the negative eigenvalues of $\rho^{T_B}$~\cite{zyczkowski1998volume,peres1996separability,horodecki2001separability}, defined as the following,
\begin{equation}
\begin{split}
 {\mathcal {N}}(\rho )=\left|\sum _{\lambda _{i}<0}\lambda _{i}\right|=\sum _{i}{\frac {|\lambda _{i}|-\lambda _{i}}{2}},
\end{split}
\end{equation}
where $\lambda_i$ are all the eigenvalues of  $\rho^{T_B}$. Negativity is a monotone and $\mathcal{N} \in [0, 0.5]$ for bipartite entanglement. Maximal entanglement states, such as Bell states, reach the value $\mathcal{N} = 0.5$. Note that the negativity is independent of which subsystem was partially transposed in Eq.~\ref{eq:pt} since $\rho^{{T_{A}}}=(\rho^{{T_{B}}})^{T}$.  

\section{Analysis of the entanglement at different $v/w$ ratio}
\begin{figure}[h]  
\centering
\includegraphics[width= 1.0\linewidth]{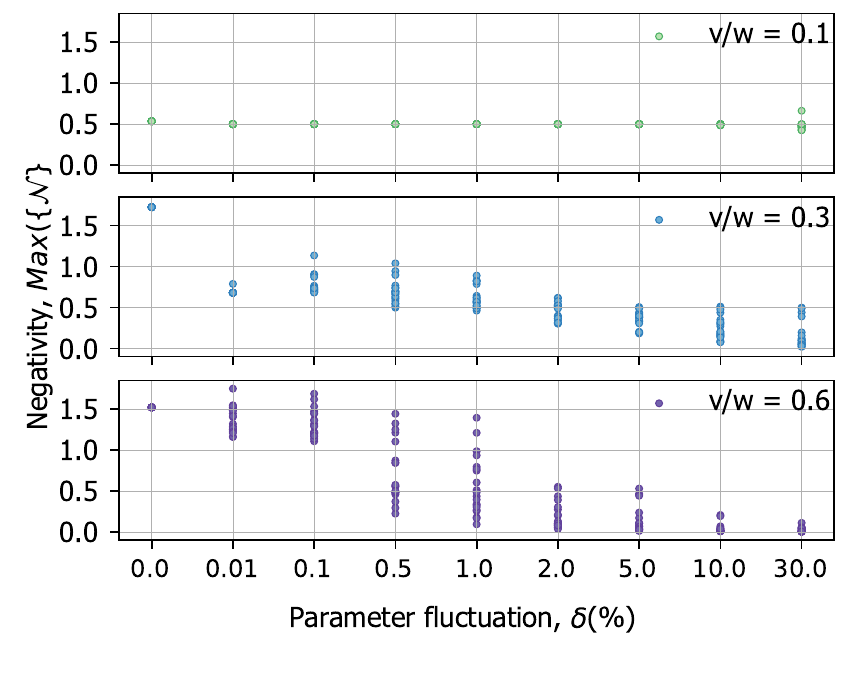}
\caption{The maximal negativity $ Max({\mathcal{N}})$ vs parameter fluctuation $\delta$ under different $v/w$ ratio. The three panels are in the topological regimes.}
\label{fig:neg_vs_v-w_ratio}
\end{figure}
In the presence of parameter fluctuation, the topological phase transition might happen at a different $v/w$ parameter value compared to fully ordered systems. Note the clear pattern difference between maximal negativity distributions in the topologically non-trivial and trivial regime as shown in Fig.~\ref{fig:neg_disorder_sample_combi}. The broadening of the distribution and decreasing of the mean value characterize the trivial phase. From the three panels with an increasing $v/w$ ratio, we find that although the systems are still in the topological regime, the negativity distribution patterns show a  similarity towards the trivial phase. This cross-over behavior characterize the change in the topological properties in the presence of disorders. The detail of this cross-over behavior is not the focus in this work. We emphasize that in order to address the target entanglement between topological modes, it is advised to stay deep in the topological regime.

\section{ Entanglement in the topological regime}
\label{app:othertypeentanglements}

Apart from the bipartite entanglement analyzed in the main text, there exist other types of entanglement in the topological regime where the negativity exceeds $0.5$, e.g., in the upper panel of Fig.~\ref{fig:neg_disorder_sample_combi}(c). These states are in the minority, which can be seen from the small amount of Hamiltonian population at these values.
The fact of $\mathcal{N} > 0.5$ means that this is an indication that something goes beyond the bipartite entanglement. When $\mathcal{N} > 0.5$, it is only considered as an indicator of outlier entanglement, instead of the entanglement measure.
We do not aim to fully understand such entanglement in this work. However, we can provide some insights into them.

\begin{figure}[t]
\centering
\includegraphics[width= 1.0\linewidth]{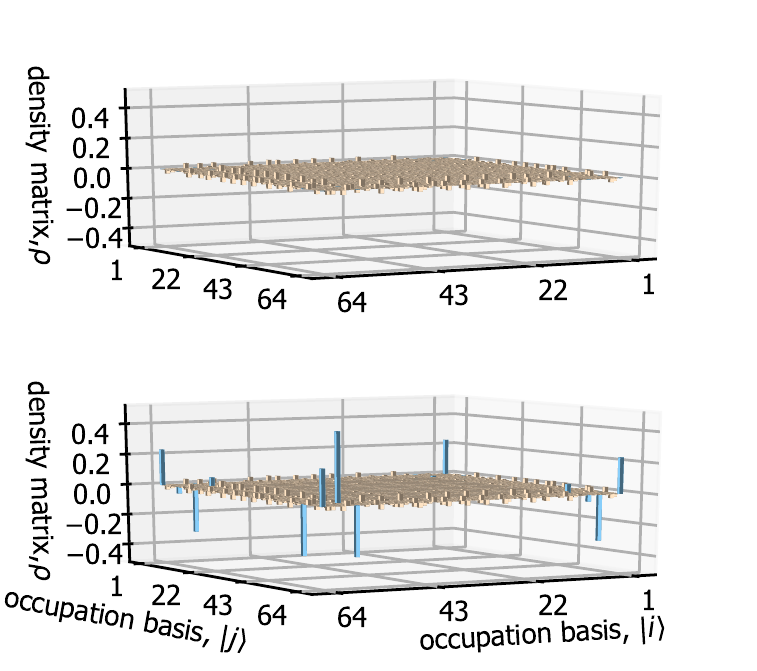}
\caption{Reduced density matrix. In the upper panel, we show the case with eliminated the edge-edge component; in the lower panel, we show the case with eliminated the bulk-bulk components. The edge-edge components are highlighted in light blue color. }
\label{fig:two_color_rho_reduced}
\end{figure}

In such outlier entanglement states, the bulk of the SSH also contributes to the entanglement. Same as in Eq.~\eqref{eq:sshbasis}, when the single excitation is at the edge sites, the basis is considered as the edge components, and vise versa.
The basis of the two SSH systems can be written as the tensor product of the two:
\begin{equation}
\ket{b} = \ket{b}_{SSH1} \otimes \ket{b}_{SSH2}.
\end{equation}
We can split the density matrix $\rho$ elements into four types according to their basis components of SSH1 and SSH2: edge-edge, edge-bulk, bulk-edge, and bulk-bulk.
We illustrate this component analysis by starting from a state in topological regime with $\mathcal{N} = 0.707$. The upper panel of Fig.~\ref{fig:two_color_rho_reduced} is the reduced density matrix with edge-edge components set to zero. It has negativity equals $ 0.671$.   Lower panel of Fig.~\ref{fig:two_color_rho_reduced} has  bulk-bulk components set to zero with negativity $ = 0.658$. The results show that the contributions from both the edge and bulk parts of the SSH are significant in the outlier entanglement states. These types of entanglement are complicated and beyond the scope of this work. 

\begin{table}[b]
\caption{\label{tab:table1}%
Energy window width and center at different $v/w$ ratios.
}
\begin{ruledtabular}
\begin{tabular}{lccr}
\textrm{$v/w$}&
\textrm{center,$\ket{e}$}&
\textrm{center, $\ket{g}$}&
\textrm{width}\\
\colrule
$0.1$ & $9.9\times 10^{-4}$ & $- 9.9\times 10^{-4}$ & $1.08\times 10^{-6}$ \\
$0.2$ & $9.6\times 10^{-4}$ & $- 9.6\times 10^{-4}$ & $4.50\times 10^{-6}$\\
$0.3$ & $9.1\times 10^{-4}$ & $- 9.1\times 10^{-4}$ & $2.87\times 10^{-6}$\\
\end{tabular}
\end{ruledtabular}
\end{table}

\section{The addressable energy window}
\label{app:energywindow}

In order to target the $\mathcal{N} = 0.5$ topological entanglement properly in the frequency space, we prepare an energy window in the following steps:

\begin{itemize}
    \item Step 1: Find out the target energies $E_{0,\ket{qubit}}$. This target energy varies for $\ket{e}$ and $\ket{g}$ qubit projection.
    \item Step 2: Find a trial energy window width $\Delta_E$. 
    \item Step 3: Prepare an energy window centered at $E_0,\ket{qubit}$ with width $\Delta_E$ for both qubit projection states.
    \item Step 4: Calculate $\mathcal{N}$ for states in this window at both qubit projections.
    \item Step 5: Note the target energy $E_0,\ket{qubit}$ shift with respect to $v/w$ ratio.
\end{itemize}

In step 1, we locate the target eigenenergies, i.e., center of the energy windows. We sample 100 Hamiltonian at each $\delta $ value in the grid of parameter fluctuation. For each Hamiltonian, at each qubit projection, we filter out the state with $\mathcal{N} = 0.5$ (rounded to 3 digits) and their energies. We collect the specific one energy with minimal absolute value $E_{sample_i,abs,min}$. We average the energy $E_{sample_i,abs,min}$ over 100 Hamiltonian samples. Thus each $\delta $ grid has an energy value. We then choose the most representative energy according to the occurrence across the fluctuation grid. The most common energy is the target energy, denoted as $E_0,\ket{qubit}$. $E_0,\ket{qubit}$ represents the mean energy of states with $\mathcal{N} = 0.5$ ($\mathcal{N}$ rounded to 3 digits) at qubit state $\ket{qubit}$.

In step 2 and 3, we prepare an energy window centered at $E_0,\ket{qubit}$ corresponding to a practical experimental setup. The trial width is chosen to be the minimum energy difference between any two states lying on the middle spectrum sector in Fig.~\ref{fig:spectrum_combi}, i.e., the orange middle sector. The challenge here is such: if the energy window is too narrow, we might miss the target state. If the energy window is too broad, we will also excite unwanted states. The choice of energy window width is empirical and can be decided from numeric studies. 

In step 4, we sample 100 Hamiltonian at each $\delta$ value in the $\delta$ grid. We calculate $\mathcal{N}$ for all states lie in the energy window. The statistics on $\mathbf{N}$ will characterize the probability of finding entangled topological states within this energy window as shown in Fig.~\ref{fig:energy_window_mean_negativity} and Fig.~\ref{fig:energy_window_prob_non_zero}.

The energy window we have used for the analysis in Fig.~\ref{fig:energy_window_mean_negativity} and Fig.~\ref{fig:energy_window_prob_non_zero} is the Tab.~\ref{tab:table1}. 
Note that the center of an energy window is relatively fixed for a specific $v/w$ ratio; the width can be adjusted to half size or double the size without losing the high targeting probability.

\nocite{*}
\bibliographystyle{unsrt}
\bibliography{ref}

\begin{thebibliography}{10}

\bibitem{wang2017topological}
Jing Wang and Shou-Cheng Zhang.
\newblock Topological states of condensed matter.
\newblock {\em Nature materials}, 16(11):1062--1067, 2017.

\bibitem{acin2018quantum}
Antonio Acin, Immanuel Bloch, Harry Buhrman, Tommaso Calarco, Christopher
  Eichler, Jens Eisert, Daniel Esteve, Nicolas Gisin, Steffen~J Glaser, Fedor
  Jelezko, et~al.
\newblock The quantum technologies roadmap: a european community view.
\newblock {\em New Journal of Physics}, 20(8):080201, 2018.

\bibitem{keimer2017physics}
B~Keimer and JE~Moore.
\newblock The physics of quantum materials.
\newblock {\em Nature Physics}, 13(11):1045--1055, 2017.

\bibitem{thouless1982quantized}
David~J Thouless, Mahito Kohmoto, M~Peter Nightingale, and Md~den Nijs.
\newblock Quantized hall conductance in a two-dimensional periodic potential.
\newblock {\em Physical review letters}, 49(6):405, 1982.

\bibitem{kane2005quantum}
Charles~L Kane and Eugene~J Mele.
\newblock Quantum spin hall effect in graphene.
\newblock {\em Physical review letters}, 95(22):226801, 2005.

\bibitem{bernevig2006quantum}
B~Andrei Bernevig and Shou-Cheng Zhang.
\newblock Quantum spin hall effect.
\newblock {\em Physical review letters}, 96(10):106802, 2006.

\bibitem{wan2011topological}
Xiangang Wan, Ari~M Turner, Ashvin Vishwanath, and Sergey~Y Savrasov.
\newblock Topological semimetal and fermi-arc surface states in the electronic
  structure of pyrochlore iridates.
\newblock {\em Physical Review B}, 83(20):205101, 2011.

\bibitem{soluyanov2015type}
Alexey~A Soluyanov, Dominik Gresch, Zhijun Wang, QuanSheng Wu, Matthias Troyer,
  Xi~Dai, and B~Andrei Bernevig.
\newblock Type-ii weyl semimetals.
\newblock {\em Nature}, 527(7579):495, 2015.

\bibitem{qi2009time}
Xiao-Liang Qi, Taylor~L Hughes, Srinivas Raghu, and Shou-Cheng Zhang.
\newblock Time-reversal-invariant topological superconductors and superfluids
  in two and three dimensions.
\newblock {\em Physical review letters}, 102(18):187001, 2009.

\bibitem{fu2010odd}
Liang Fu and Erez Berg.
\newblock Odd-parity topological superconductors: theory and application to cu
  x bi 2 se 3.
\newblock {\em Physical review letters}, 105(9):097001, 2010.

\bibitem{sasaki2011topological}
Satoshi Sasaki, M~Kriener, Kouji Segawa, Keiji Yada, Yukio Tanaka, Masatoshi
  Sato, and Yoichi Ando.
\newblock Topological superconductivity in cu x bi 2 se 3.
\newblock {\em Physical review letters}, 107(21):217001, 2011.

\bibitem{qi2010chiral}
Xiao-Liang Qi, Taylor~L Hughes, and Shou-Cheng Zhang.
\newblock Chiral topological superconductor from the quantum hall state.
\newblock {\em Physical Review B}, 82(18):184516, 2010.

\bibitem{vzutic2004spintronics}
Igor {\v{Z}}uti{\'c}, Jaroslav Fabian, and S~Das Sarma.
\newblock Spintronics: Fundamentals and applications.
\newblock {\em Reviews of modern physics}, 76(2):323, 2004.

\bibitem{aasen2016milestones}
David Aasen, Michael Hell, Ryan~V Mishmash, Andrew Higginbotham, Jeroen Danon,
  Martin Leijnse, Thomas~S Jespersen, Joshua~A Folk, Charles~M Marcus, Karsten
  Flensberg, et~al.
\newblock Milestones toward majorana-based quantum computing.
\newblock {\em Physical Review X}, 6(3):031016, 2016.

\bibitem{bernevig2013topological}
B~Andrei Bernevig and Taylor~L Hughes.
\newblock {\em Topological insulators and topological superconductors}.
\newblock Princeton university press, 2013.

\bibitem{susstrunk2015observation}
Roman S{\"u}sstrunk and Sebastian~D Huber.
\newblock Observation of phononic helical edge states in a mechanical
  topological insulator.
\newblock {\em Science}, 349(6243):47--50, 2015.

\bibitem{susstrunk2016classification}
Roman Susstrunk and Sebastian~D Huber.
\newblock Classification of topological phonons in linear mechanical
  metamaterials.
\newblock {\em Proceedings of the National Academy of Sciences},
  113(33):E4767--E4775, 2016.

\bibitem{kane2014topological}
CL~Kane and TC~Lubensky.
\newblock Topological boundary modes in isostatic lattices.
\newblock {\em Nature Physics}, 10(1):39, 2014.

\bibitem{paulose2015topological}
Jayson Paulose, Bryan Gin-ge Chen, and Vincenzo Vitelli.
\newblock Topological modes bound to dislocations in mechanical metamaterials.
\newblock {\em Nature Physics}, 11(2):153, 2015.

\bibitem{chen2014nonlinear}
Bryan Gin-ge Chen, Nitin Upadhyaya, and Vincenzo Vitelli.
\newblock Nonlinear conduction via solitons in a topological mechanical
  insulator.
\newblock {\em Proceedings of the National Academy of Sciences},
  111(36):13004--13009, 2014.

\bibitem{chen2016topological}
Bryan Gin-ge Chen, Bin Liu, Arthur~A Evans, Jayson Paulose, Itai Cohen,
  Vincenzo Vitelli, and CD~Santangelo.
\newblock Topological mechanics of origami and kirigami.
\newblock {\em Physical review letters}, 116(13):135501, 2016.

\bibitem{nash2015topological}
Lisa~M Nash, Dustin Kleckner, Alismari Read, Vincenzo Vitelli, Ari~M Turner,
  and William~TM Irvine.
\newblock Topological mechanics of gyroscopic metamaterials.
\newblock {\em Proceedings of the National Academy of Sciences},
  112(47):14495--14500, 2015.

\bibitem{serra2018observation}
Marc Serra-Garcia, Valerio Peri, Roman S{\"u}sstrunk, Osama~R Bilal, Tom
  Larsen, Luis~Guillermo Villanueva, and Sebastian~D Huber.
\newblock Observation of a phononic quadrupole topological insulator.
\newblock {\em Nature}, 555(7696):342, 2018.

\bibitem{imhof2018topolectrical}
Stefan Imhof, Christian Berger, Florian Bayer, Johannes Brehm, Laurens~W
  Molenkamp, Tobias Kiessling, Frank Schindler, Ching~Hua Lee, Martin Greiter,
  Titus Neupert, et~al.
\newblock Topolectrical-circuit realization of topological corner modes.
\newblock {\em Nature Physics}, 14(9):925--929, 2018.

\bibitem{kollar2019hyperbolic}
Alicia~J Koll{\'a}r, Mattias Fitzpatrick, and Andrew~A Houck.
\newblock Hyperbolic lattices in circuit quantum electrodynamics.
\newblock {\em Nature}, 571(7763):45--50, 2019.

\bibitem{kim2021quantum}
Eunjong Kim, Xueyue Zhang, Vinicius~S Ferreira, Jash Banker, Joseph~K Iverson,
  Alp Sipahigil, Miguel Bello, Alejandro Gonz{\'a}lez-Tudela, Mohammad
  Mirhosseini, and Oskar Painter.
\newblock Quantum electrodynamics in a topological waveguide.
\newblock {\em Physical Review X}, 11(1):011015, 2021.

\bibitem{su1979solitons}
W\_P Su, JR~Schrieffer, and Ao~J Heeger.
\newblock Solitons in polyacetylene.
\newblock {\em Physical review letters}, 42(25):1698, 1979.

\bibitem{blais2004cavity}
Alexandre Blais, Ren-Shou Huang, Andreas Wallraff, Steven~M Girvin, and R~Jun
  Schoelkopf.
\newblock Cavity quantum electrodynamics for superconducting electrical
  circuits: An architecture for quantum computation.
\newblock {\em Physical Review A}, 69(6):062320, 2004.

\bibitem{koch2007charge}
Jens Koch, M~Yu Terri, Jay Gambetta, Andrew~A Houck, David~I Schuster, Johannes
  Majer, Alexandre Blais, Michel~H Devoret, Steven~M Girvin, and Robert~J
  Schoelkopf.
\newblock Charge-insensitive qubit design derived from the cooper pair box.
\newblock {\em Physical Review A}, 76(4):042319, 2007.

\bibitem{wendin2017quantum}
G{\"o}ran Wendin.
\newblock Quantum information processing with superconducting circuits: a
  review.
\newblock {\em Reports on Progress in Physics}, 80(10):106001, 2017.

\bibitem{frey2012dipole}
T~Frey, PJ~Leek, M~Beck, Alexandre Blais, Thomas Ihn, Klaus Ensslin, and
  Andreas Wallraff.
\newblock Dipole coupling of a double quantum dot to a microwave resonator.
\newblock {\em Physical Review Letters}, 108(4):046807, 2012.

\bibitem{houck2008controlling}
AA~Houck, JA~Schreier, BR~Johnson, JM~Chow, Jens Koch, JM~Gambetta,
  DI~Schuster, L~Frunzio, MH~Devoret, SM~Girvin, et~al.
\newblock Controlling the spontaneous emission of a superconducting transmon
  qubit.
\newblock {\em Physical review letters}, 101(8):080502, 2008.

\bibitem{scigliuzzo2021extensible}
Marco Scigliuzzo, Giuseppe Calaj{\`o}, Francesco Ciccarello, Daniel~Perez
  Lozano, Andreas Bengtsson, Pasquale Scarlino, Andreas Wallraff, Darrick
  Chang, Per Delsing, and Simone Gasparinetti.
\newblock Extensible quantum simulation architecture based on atom-photon bound
  states in an array of high-impedance resonators.
\newblock {\em arXiv preprint arXiv:2107.06852}, 2021.

\bibitem{scarlino2019all}
Pasquale Scarlino, David~J Van~Woerkom, Anna Stockklauser, Jonne~V Koski,
  Michele~C Collodo, Simone Gasparinetti, Christian Reichl, Werner Wegscheider,
  Thomas Ihn, Klaus Ensslin, et~al.
\newblock All-microwave control and dispersive readout of gate-defined quantum
  dot qubits in circuit quantum electrodynamics.
\newblock {\em Physical review letters}, 122(20):206802, 2019.

\bibitem{de2010hybrid}
Silvano De~Franceschi, Leo Kouwenhoven, Christian Sch{\"o}nenberger, and
  Wolfgang Wernsdorfer.
\newblock Hybrid superconductor--quantum dot devices.
\newblock {\em Nature nanotechnology}, 5(10):703--711, 2010.

\bibitem{delbecq2011coupling}
MR~Delbecq, Vivien Schmitt, FD~Parmentier, Nicolas Roch, JJ~Viennot, Gwendal
  F{\`e}ve, Benjamin Huard, Christophe Mora, Audrey Cottet, and Takis Kontos.
\newblock Coupling a quantum dot, fermionic leads, and a microwave cavity on a
  chip.
\newblock {\em Physical Review Letters}, 107(25):256804, 2011.

\bibitem{hensgens2017quantum}
Toivo Hensgens, Takafumi Fujita, Laurens Janssen, Xiao Li, CJ~Van~Diepen,
  Christian Reichl, Werner Wegscheider, Sankar Das~Sarma, and Lieven~MK
  Vandersypen.
\newblock Quantum simulation of a fermi--hubbard model using a semiconductor
  quantum dot array.
\newblock {\em Nature}, 548(7665):70--73, 2017.

\bibitem{nowack2007coherent}
Katja~C Nowack, FHL Koppens, Yu~V Nazarov, and LMK Vandersypen.
\newblock Coherent control of a single electron spin with electric fields.
\newblock {\em Science}, 318(5855):1430--1433, 2007.

\bibitem{hendrickx2021four}
Nico~W Hendrickx, William~IL Lawrie, Maximilian Russ, Floor van Riggelen,
  Sander~L de~Snoo, Raymond~N Schouten, Amir Sammak, Giordano Scappucci, and
  Menno Veldhorst.
\newblock A four-qubit germanium quantum processor.
\newblock {\em Nature}, 591(7851):580--585, 2021.

\bibitem{zyczkowski1998volume}
Karol {\.Z}yczkowski, Pawe{\l} Horodecki, Anna Sanpera, and Maciej Lewenstein.
\newblock Volume of the set of separable states.
\newblock {\em Physical Review A}, 58(2):883, 1998.

\bibitem{peres1996separability}
Asher Peres.
\newblock Separability criterion for density matrices.
\newblock {\em Physical Review Letters}, 77(8):1413, 1996.

\bibitem{horodecki2001separability}
Micha{\l} Horodecki, Pawe{\l} Horodecki, and Ryszard Horodecki.
\newblock Separability of n-particle mixed states: necessary and sufficient
  conditions in terms of linear maps.
\newblock {\em Physics Letters A}, 283(1-2):1--7, 2001.

\bibitem{besse2020realizing}
Jean-Claude Besse, Kevin Reuer, Michele~C Collodo, Arne Wulff, Lucien Wernli,
  Adrian Copetudo, Daniel Malz, Paul Magnard, Abdulkadir Akin, Mihai Gabureac,
  et~al.
\newblock Realizing a deterministic source of multipartite-entangled photonic
  qubits.
\newblock {\em Nature communications}, 11(1):1--6, 2020.

\bibitem{eichler2012characterizing}
Christopher Eichler, Deniz Bozyigit, and Andreas Wallraff.
\newblock Characterizing quantum microwave radiation and its entanglement with
  superconducting qubits using linear detectors.
\newblock {\em Physical Review A}, 86(3):032106, 2012.

\bibitem{TopoEntanglementStabilization}
Guliuxin Jin and Eliska Greplova.
\newblock Topological entanglement stabilization.
\newblock \url{https://gitlab.com/QMAI/papers/topoentanglementstabilization},
  2022.

\bibitem{owens2018quarter}
Clai Owens, Aman LaChapelle, Brendan Saxberg, Brandon~M Anderson, Ruichao Ma,
  Jonathan Simon, and David~I Schuster.
\newblock Quarter-flux hofstadter lattice in a qubit-compatible microwave
  cavity array.
\newblock {\em Physical Review A}, 97(1):013818, 2018.

\bibitem{uhlmann1976transition}
Armin Uhlmann.
\newblock The "transition probability" in the state space of a*-algebra.
\newblock {\em Reports on Mathematical Physics}, 9(2):273--279, 1976.

\bibitem{alberti1983note}
Peter~M Alberti.
\newblock A note on the transition probability over c*-algebras.
\newblock {\em Letters in Mathematical Physics}, 7(1):25--32, 1983.

\bibitem{alberti1983stochastic}
Peter~M Alberti and Armin Uhlmann.
\newblock Stochastic linear maps and transition probability.
\newblock {\em Letters in Mathematical Physics}, 7(2):107--112, 1983.

\end{thebibliography}
\end{document}